\documentclass[aps,prl,twocolumn,groupedaddress]{revtex4}
\usepackage{graphics}

\begin{document}

\title{Renormalization scheme for a multi-qubit-network}
\author{Alexandra Olaya-Castro}
\email{a.olaya@physics.ox.ac.uk}
\author{Chiu Fan \surname{Lee}}
\author{Neil F. Johnson}

\affiliation{Centre for Quantum Computation and Physics
Department, University of Oxford, Parks Road, OX1 3PU, U.K.}


\begin{abstract}
We present a renormalization scheme which simplifies the dynamics
of an important class of interacting multi-qubit systems. We show
that a wide class of $M+1$ qubit systems can be reduced to an
equivalent $n+1$ qubit system with $n\geq 2$, for {\em any} $M$.
Our renormalization scheme faithfully reproduces the overall
dynamics of the original system including the entanglement
properties. In addition to its direct application to atom-cavity
and nanostructure systems, the formalism offers insight into a
variety of situations ranging from decoherence due to a spin-bath
with its own internal entanglement, through to energy transfer processes in
organic systems such as biological photosynthetic units.
\end{abstract}

\pacs{}

\maketitle

Many-body problems are very difficult, if not impossible, to solve
exactly. Any simplifications are therefore of great potential
importance -- not only because of practical applications but also
because of a basic theoretical interest. Such simplifications
usually result if some exact or approximate symmetry can be
identified in the underlying Hamiltonian (e.g. Ref.
\cite{calogero}). In terms of practical applications, most exact
results to date have concerned systems where all the interacting
objects (e.g. particles) are indistinguishable (e.g. Ref. \cite{laughlin}).
This is understandable, since systems such as
many-electron gases have been of great experimental interest over
the past few decades. However given the current levels of activity
in the field of quantum information processing, there is a clear
desire to develop such theoretical results for multi-qubit
systems. In particular, it would be highly desirable to obtain
exact or approximate results for the generic situation in which a
collection of qubits interacts with an auxiliary system such as a
cavity mode \cite{cavity_mode} or a central spin
\cite{central_spin}.
\begin{figure}[b]
\resizebox{6cm}{!}{\includegraphics{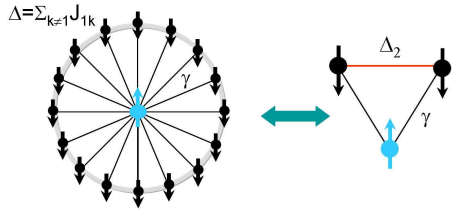}} \caption {
Schematic representation of the renormalization scheme in which an
interacting $M+1$ multi-qubit with one excitation can be reduced
to an equivalent $n+1$ qubit system with one excitation, where
$n\geq 2$. For example, the excitation can correspond to having a
central qubit in its excited state (spin up) while the outer
qubits are in their ground state (spin down). A wide range of
interactions are possible between outer qubits, including
nearest-neighbor, dipole-dipole, and pairwise interactions between
any pair in the ring.} \label{fig:equiv}
\end{figure}
\begin{figure}
\begin{center}
\resizebox{3.5cm}{!}{\includegraphics*{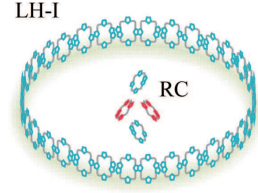}}
\caption{Schematic diagram of the light-harvesting complex LH-I
and reaction centre RC structure in purple bacteria. For details
see Ref. \cite{hu97}. This photosynthetic unit closely resembles
our model system in Fig. \ref{fig:equiv}.} \label{fig:lhrc}
\end{center}
\end{figure}

Here we show that a system of $M+1$ qubits whose interactions
resemble a spin-star configuration (see Fig. \ref{fig:equiv}) can
be mapped onto an equivalent $n+1$ interacting system with $n\geq
2$, preserving the dynamics of the central qubit {\em and} the
quantum correlations of the original system. This setup can be
realized using multi-atom-cavity systems, and could also be
engineered from a collection of quantum dots in an optical cavity.
It also mimics naturally-occuring photosynthetic complexes (see
Fig. \ref{fig:lhrc}) which are of fundamental importance in
nature. Moreover, we will show that it yields insight into the
decoherence properties of spin-baths possessing internal
entanglement.

The key physical feature which underpins this equivalence, is a
particular non-trivial symmetry in the interactions experienced by
all outer qubits. This symmetry allows us to describe the system's
dynamics in terms of two time-dependent variables, associated with
the outer qubits and central qubit (or cavity mode) respectively.
We consider $M$ identical qubits which interact among themselves
and with a central qubit via the following interaction Hamiltonian
(c.f. Fig. 1):
\begin{eqnarray}
H_I & = &\sum_{j=1}^M\gamma_{jC}
\{\sigma_C^{+} \sigma_j^{-}+\sigma_j^{+}\sigma_C^{-}\}  \nonumber \\
&  & + \sum_j\sum_{k \neq j}J_{jk}\{\sigma_j^{+}
\sigma_k^{-}+\sigma_k^{+}\sigma_j^{-}\}
\end{eqnarray} where $\sigma_j^{+(-)}$ is the usual raising (lowering) operator
for $j$'th outer qubit, or the central qubit with $j\equiv C$. The
formalism and results in this paper apply to a wide range of
possible two-body interactions $J_{jk}$, e.g. short-range (i.e.
nearest-neighbor), long-range (i.e. pairwise, between any pair in
the ring) and dipole-dipole. The Hamiltonian $H_I$ preserves the
number of excitations, i.e $[H_I,{\mathcal N}]=0$ with ${\mathcal
N}=\sum_{j=1}^M\sigma_j^{+} \sigma_j^{-}+\sigma_C^{+}
\sigma_C^{-}$. We focus first on the dynamics within the
single-excitation subspace. A basis is given by states in which
one qubit is excited and the rest are in their ground state, i.e.
{\small$\{|q_1\rangle|0_C\rangle, |q_j\rangle|0_C\rangle,...
|q_M\rangle|0_C\rangle, |0_B\rangle|1_C\rangle \}$} with
\begin{eqnarray}
|q_j\rangle& = &  |0_1,0_2\cdots 1_j\cdots 0_M\rangle \\ \nonumber
|0_B\rangle& =& |0_1,0_2\cdots 0_j\cdots 0_M\rangle
\end{eqnarray}
with $j=1,\dots M$. The state of this $M$-plus-central qubit system
is given by the unitary evolution associated with $H_I$:
\begin{eqnarray}
|\Psi(t)\rangle = |1_B\rangle|0_C\rangle + b_{C}(t)|0_B\rangle|1_C\rangle
\label{eq:state}
\end{eqnarray}
with $|1_B\rangle=\sum_{j=1}^{M}b_j(t)|q_j\rangle$ and
$\sum_{j=1}^{M}|b_j(t)|^2 +|b_C(t)|^2=1$. We have written the
state in this way in order to emphasize the collective behaviour
of the outer qubits -- indeed, as a reference to one of the
potentially important applications of this work, we will
frequently use the term {\em bath} to denote this collection of
outer qubits. The system's state satisfies the Schr\"odinger
equation $d|\Psi(t)\rangle/dt=-iH_I|\Psi(t)\rangle$, which leads
to a set of first-order coupled differential equations for the
complex amplitudes $b_j(t)$ and $b_C(t)$. Due to a collective
symmetry in the effective interaction experienced by all outer
qubits, it is possible to describe the system's dynamics in terms
of two time-dependent variables $b_c(t)$ and
$B(t)=\sum_{j=1}^{M}b_j(t)$ associated with the central qubit and
all outer qubits respectively. The nature of this symmetry is as
follows: Let us assume that each one of the outer qubits interacts
with the center via an identical coupling $\gamma_{jC}\equiv
\gamma$. Because of this, the strength of the effective
interaction between a qubit $j$ and the rest of the outer qubits
is captured by a parameter $\Delta_j = \sum_{k \neq j} J_{jk}$,
i.e. the sum of all the coupling strengths between qubit $j$ and
any other outer qubit. When $\Delta_j$ is identical for all
qubits, i.e. $\Delta_j \equiv \Delta$, as it is for example in the
case of nearest-neighbour or pairwise interactions, the system's
dynamics can be described in terms of $b_c(t)$ and $B(t)$. Note
that this does not imply that the pair couplings $J_{jk}$ need to
be identical for all possible pairs. Instead it implies that the
collective interaction experienced by one of the outer qubits, due
to the rest, should be identical for all qubits. Hence the
symmetry is such that the  interaction experienced for each qubit
(central or outer) is of a collective nature. Given this symmetry,
$b_c(t)$ and $B(t)$ satisfy the following set of differential
equations:
$\ddot b_C(t) +i\Delta \dot b_C(t) + 4M\gamma^2 b_C(t) = 0$ and
$\dot B(t) \: + \: i\Delta B(t) = -i M \gamma b_C(t)$,
with solutions of the form $b_C(t) = f_C(t)B(0)+ g_C(t)b_C(0)$ and
$B(t) = f_B(t)B(0)+ g_B(t)b_C(0)$
where \begin{eqnarray}
f_C(t) &=&-ie^{-i\Delta t/2}2\gamma u(t)/\Omega\nonumber\\ \nonumber
g_C(t) &=&e^{-i\Delta t/2}[(i\Delta/\Omega)u(t) +v(t)]\\ \nonumber
f_B(t) &=&e^{-i\Delta t/2}[(i\Delta/\Omega)u(t) -v(t)]\\
g_B(t) &=&-ie^{-i\Delta t/2}2M\gamma u(t)/\Omega
\end{eqnarray}
with $u(t)=\rm{sin}(\Omega t/2)$, $v(t)=\rm{cos}(\Omega t/2)$, and
\begin{eqnarray}
\Omega=\Omega_M(\gamma,\Delta)=\sqrt{4M\gamma^2 +\Delta^2}
\label{eq:omega}
\end{eqnarray}
is a frequency that captures the collective character of the
interactions in the system. Note that $2\sqrt{M}\gamma$ captures
the collective features of the interaction between the outer and
central qubits, while $\Delta$ captures the collective interaction
between one outer qubit and the remaining $M-1$ qubits. The
relevant expectation values for the dynamics of the central qubit
are determined by the reduced density operator $\rho_C(t)={\rm
Tr}_{outer}\{|\Psi(t)\rangle\langle\Psi(t)|\}$. For an initial
state of the same form as Eq.(\ref{eq:state}) we have
\begin{eqnarray}
\rho_C(t)=|b_C(t)|^2|0_C\rangle\langle 0_C| +
(1-|b_C(t)|^2)|1_C\rangle\langle 1_C| \label{eq:cq} \ .
\end{eqnarray}
Hence all the physical properties of the central qubit are
determined by $b_C(t)$ and the relevant expectation values can be
calculated, e.g. $\langle \sigma^z_{c}\rangle=2|b_C(t)|^2-1$.

The system's dynamics are characterized by two effective
interaction strengths $\gamma$ and $\Delta$. This suggests that
the entanglement properties should also be describable in terms of
two contributions \cite{dawson05}: one corresponding to the
entanglement among all outer qubits, which we call ${\mathcal
E}_{B}$, and another corresponding to the entanglement between the
central qubit and the rest, which we call ${\mathcal E}_{BC}$. The
time-dependent versions of these quantities are given by
\begin{eqnarray}
{\mathcal E}_{B}(t)=|-1+|B(t)|^2+|b_C(t)|^2|
\label{eq:ent}
\end{eqnarray}
and
\begin{eqnarray}
{\mathcal E}_{BC}(t)=4|b_C(t)|^2(1-|b_C(t)|^2)
\label{eq:ebc}
\end{eqnarray}
We now take advantage of some known results for $W-$states since
these are the states of interest in our multi-qubit system (see
Eq.~\ref{eq:state}). In particular, for any partition in the
system the entanglement is entirely composed of pairwise
contributions \cite{dawson05} that can be quantified by the
concurrence \cite{wootters98}. The concurrence of the reduced
state of two qubits in our system has the form
$C_{jk}=2|b_j(t)b_k(t)^*|\leq 1$. Hence a measure of the total
entanglement among the outer qubits is $E_B=\sum_{\langle
j,k\rangle}C_{jk}$, while the total entanglement between the
central and outer qubits is $E_{BC}=\sum_{j=1}^M C_{jC}$. We now
demonstrate that ${\mathcal E}_{B}(t)$ and ${\mathcal E}_{BC}(t)$
are lower bounds of $E_B$ and $E_{BC}$, respectively i.e.
${\mathcal E}_B(t)\leq E_B$, and ${\mathcal E}_{BC}(t)\leq
E_{BC}$, and hence they can be used to quantify the entanglement
properties of our system. We express the complex amplitudes
$b_j(t)$ as $b_j(t)=|b_j(t)|e^{i\theta_k}$ hence
\begin{eqnarray*}
|B(t)|^2 & = & \sum_{j=1}^M\sum_{k=1}^M |b_j(t)b_k(t)^*|e^{i\alpha_{jk}}\;\;
{\mbox {\small with $\alpha_{jk}=\theta_j-\theta_k$}}
\\& = & \sum_{j=1}^M|b_j(t)|^2+\sum_{\langle j,k\rangle}C_{jk}{\rm cos}(\alpha_{jk})
\end{eqnarray*}
Using $1=\sum_{j=1}^M|b_j(t)|^2+|b_C(t)|^2$, we have
\[
{\mathcal E}_B(t) =
\Bigg|\sum_{\langle j,k\rangle}C_{jk}{\rm cos}(\alpha_{jk}) \Bigg|
 \leq \sum_{\langle j,k\rangle}C_{jk}|{\rm cos}(\alpha_{jk})|
 \leq \sum_{\langle j,k\rangle}C_{jk}
\]
Knowing that $C_{jC}\leq 1$, we then have that
$E_{BC}\geq \sum_{j=1}^M C_{jC}^2=4|b_C(t)|^2(1-|b_C(t)|^2)={\mathcal E}_{BC}(t)$
which completes our proof.

An important observation from the above relations is that the
central qubit dynamics (Eq.(\ref{eq:cq})) and  the central-outer
qubit entanglement (Eq. (\ref{eq:ebc})) are both completely
determined by the probability of having the excitation on the
central qubit, i.e. $|b_C(t)|^2$, while the intra-bath
entanglement ${\mathcal E}_B(t)$ depends on both $|b_C(t)|^2$ and
$|B_0(t)|^2$. Therefore $|b_C(t)|^2$ and ${\mathcal E}_B(t)$ are
the relevant quantities to characterize the dynamics of the
$M+1$-qubit system. Henceforth we shall refer to the quantities as
$P_M(t)=|b_C(t)|^2$ and ${E}_M(t)={\mathcal E}_B(t)$ where the
subscript $M$ indicates that these quantities correspond to the
$M+1$ system.

The above considerations lead to the following dynamical equivalence, which we shall now discuss:
{\em  A system of $M+1$ qubits with interactions forming a spin-star configuration and
characterized by two parameters $\gamma$ and $\Delta$, is dynamically equivalent
to a system of $n+1$ qubits in a similar configuration and characterized
by $\Delta_{n}$ and $\gamma_{n}$} given that their
collective frequencies are identical
$4M\gamma^2 +\Delta^2=4n\gamma_{n}^2 +\Delta_{n}^2$.
Since $\gamma$ and $\Delta$ are independent of each other, we can fix
$\gamma$ in the equivalent and original systems, i.e $\gamma=\gamma_{n}$, such that
\begin{eqnarray}
\Delta_{n}^2=4(M-n)\gamma + \Delta^2 \label{eq:deq}
\end{eqnarray}
The above statements imply that we can find transformations
between the dynamical quantities of the original and equivalent
systems, i.e. $P_M(t)={\mathcal F}(P_{n}(t))$ and
${E}_M(t)={\mathcal F}({E}_{n}(t))$. These
transformations are:
\begin{eqnarray}
\frac{P_M(t)-P_M(0)}{\alpha_M} & = &
\frac{P_{n}(t)-P_{n}(0)}{\alpha_n} \nonumber \\
\frac{{E}_M(t)-{E}_M(0)}{\beta_M} & = & \frac{{
E}_{n}(t)-{E}_{n}(0)}{\beta_n} \label{eq:eqv}
\end{eqnarray}
with $\alpha_n=|B_n(0)|^2-n P_n(0)$ and
$\beta_n=n(n-1)P_n(0)+(n-1)|B_n(0)|^2$. Notice that $\beta_n$ sets
the minimum $n$ in the above relations to two, i.e. $n_{min}=2$.
Therefore {\em in order to preserve the entanglement
properties of the original system, the equivalent system should
have at least two outer qubits.} Hence as  far as the dynamics of
the central qubit and the overall entanglement properties are
concerned, it is possible to map an $M+1$ qubit system onto a
system of $2+1$ interacting qubits (c.f. Fig.~1).
As an example, consider the case where the excitation is
initially on the central spin. The time-evolution for $P_{M}(t)/M$
and ${E}_{M}(t)$ in an original system with $M=10$ and the
equivalent system $M=2$, are shown in Fig.~2. They are described
by the following relations:
\begin{eqnarray}
P_M(t)& = &\left [1- \frac{M}{2} \right ]
+\frac{M}{2}P_2(t)\nonumber \\
{E}_M(t)& = & \frac{M(M-1)}{2}{\mathcal E}_2(t) \ .
\label{eq:eqv1}
\end{eqnarray}
\begin{figure}[t]
\resizebox{8cm}{!}{\includegraphics{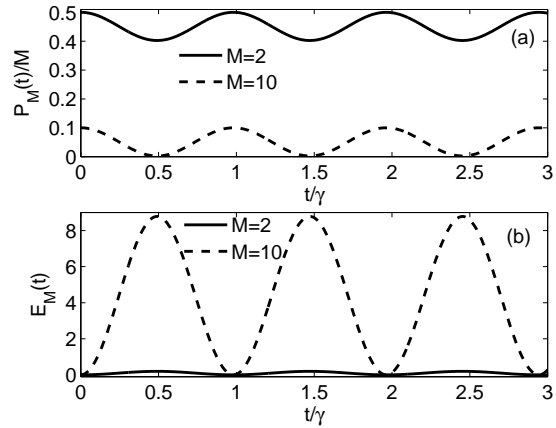}}
\caption[Central qubit dynamics and intra-bath entanglement
properties for original and equivalent systems. The excitation is
initially on the central qubit.]{(a) Dynamics of central qubit, and (b)
intra-bath entanglement properties for original and equivalent
systems when the excitation is initially on the central qubit,
i.e. $b_c(0)=1$. Nearest-neighbour interactions have been
assumed such that $\Delta=J$ and $\Omega=\sqrt{40\gamma + J}$. We
take $J=\gamma$.} \label{fig:bc1}
\end{figure}
We now discuss the effect of dissipation in the system using the
quantum jump approach\cite{qja}. We assume identical decay-rates
for the outer qubits $\Gamma$, but these can be different from the
decay-rate $\kappa$ of the central qubit. The non-unitary dynamics
conditioned on `no-loss' of excitation is given by
$H_{cond}=H_I-i\Gamma\sum_{j=1,M}\sigma_j^{+}\sigma_j^{-}
+i\kappa\sigma_C^{+} \sigma_C^{-}$. The unnormalized state of the
system is $|\Psi_{cond}(t)\rangle={\rm
exp}[-iH_{cond}t]|\Psi(0)\rangle$ which can be expressed  in the
same form as in Eq.(\ref{eq:state}). Hence, we can again describe
the system's dynamics in terms of (normalized versions of)
$b_C(t)$ and $B(t)$. In this case, one obtains:
\begin{eqnarray}
f_C(t) &=&-ie^{-Xt/2}2\gamma u(t)/\Omega\nonumber\\ \nonumber
g_C(t) &=&e^{-Xt/2}\left[\frac{(\delta +i\Delta)}{\Omega}u(t)
+v(t)\right]\\ \nonumber
f_B(t) &=&e^{-Xt/2}\left[\frac{(\delta +i\Delta)}{\Omega}u(t)-v(t)\right]\\
g_B(t) &=&-ie^{-Xt/2}2M\gamma u(t)/\Omega
\end{eqnarray}
where $X=\kappa +\Gamma + i\Delta$,
$\delta=\Gamma-\kappa$ is the effective dissipation coefficient, and $u(t)$ and $v(t)$ are
defined as before. The collective frequency is now given by
\begin{eqnarray}
\Omega=\Omega_M(\gamma, \Delta, \delta)=\sqrt{4M\gamma^2
-(\delta+i\Delta)^2} \ .
\end{eqnarray}
With the above equations, one can obtain a similar renormalization
scheme to the one discussed earlier.

The above formalism can be immediately applied to model a quantum
spin-bath with intra-environmental coupling. The spin-bath model
has typically been employed in the literature to describe the
system-bath interaction at low temperatures.  In order to achieve
tractable approaches, interactions among the bath spins have
usually been neglected. However, recent works have started to
explore whether the interaction among bath spins might indeed have
significant knock-on effects on the dynamics of the central qubit
\cite{tessieri02,dawson05}. These studies suggest that the
intra-bath interactions can suppress decoherence of the central
spin. It has also been argued that such an effect is due to the
fact that the intra-bath entanglement limits the spin-bath
entanglement, and hence the decoherence of the central spin
\cite{dawson05}. These features can be easily understood within
our approach, as follows. The intra-bath interaction is
represented by the effective coupling $\Delta$, which affects the
dynamics of the central spin through the collective effective
frequency $\Omega$. The relation between the spin-bath
entanglement (${\mathcal E}_{BC}(t)$) and the intra-bath
entanglement (${\mathcal E}_B(t)$) can be seen explicitly in
Eq.(\ref{eq:ent}) where it is clear that ${\mathcal E}_{B}(t)$
limits $|b_C(t)|^2$, thereby limiting the spin-bath entanglement.
Most importantly, our approach indicates that such complex
many-body features can be simulated by a simple system of only
{\em three} interacting qubits.

Our theoretical formalism can also be used to investigate the
excitation transfer between a light-harvesting LH-I complex and
the reaction centre RC in photosynthetic bacteria (see Fig.
\ref{fig:lhrc} and Ref. \cite{hu97}). The LH-I ring is made up of
32 donor units -- each of these \cite{hu97}, in addition to the
RC, can be treated as a two-level system to a good approximation.
The prevalent interaction between donors is given by an induced
dipole-dipole coupling which can be approximated as $J_{jk}\simeq
J/r_{jk}^3$ where ${\boldmath \rm r}_{jk}$ is the relative
position vector between the outer qubits $j$ and $k$. To a good
approximation, all the induced dipole moments can be taken as
identical and lying perpendicular to the plane containing the
outer qubits, yielding $\Delta=J\sum^M_{k=2}(1/r_{1k})^3$. Hence,
as far as the dynamics of the center is concerned, this
complicated 32-donor ring can be accurately represented by {\em
two} donors. We do not pursue this any further here, but leave it
as an interesting consequence of the present theoretical study.

We now discuss the extension of these results to larger numbers of
excitations in the multi-qubit system. Since the central qubit
behaves as a spin-$\frac{1}{2}$ particle, only two dimensions of
the bath's state-space are required to expand the pure state in
its Schmidt decomposition. For any number of excitations $N\leq
M$, we can express the system's state in an analogous way to
Eq.(\ref{eq:state})\begin{eqnarray}
|\Psi(t)\rangle=|N_B\rangle|0_C\rangle + |N-1_B\rangle|1_C\rangle
\end{eqnarray}
where $|A_B\rangle$ is a time-dependent superposition of ${M
\choose A}$ states, each one having $A$ excitations. We can
therefore justifiably claim that the dynamics of a system with
$N\leq M$ excitations is analogous to the case of $M+1-N$
excitations. A particular case is $N=M$ whose solutions are
analogous to the case of single excitation case we have discussed.
This can be seen in Fig. \ref{fig:equiv} but now interpreting the
presence of an excitation as a spin-down. Therefore, {\em a system
with $M+1$ qubits and $M$ excitations can be mapped on to a
$2+1$ system with $2$ excitations}. The complexity of the dynamics
is, however, highly nontrivial for $M-N>0$. Although we don't
currently have a full solution, insight into this problem
can be gained by analyzing the effect of $H_I$ on
$|\Psi(t)\rangle$ in detail. In particular, we re-write
$H_I=V_B+V_{BC}$ where $V_B$ is the interaction between central
qubit and the outer qubits, and $V_{BC}$ represents the
interaction among outer qubits. $V_{BC}$ induces
transitions from $|N_B\rangle|0_C\rangle$ to
$|N-1_B\rangle|1_C\rangle$ while $V_B$ just produces internal
transitions in each $|A_B\rangle$, since it preserves the internal
number of excitations in the outer qubits. Hence we conjecture
that for any $N \leq M$, the dynamics can be described in terms of
two collective variables: one associated with
$|N_B\rangle|0_C\rangle$ and the other with
$|N-1_B\rangle|1_C\rangle$. The open question for future study
then becomes: What is the {\em minimum} number of qubits needed
to represent such collective properties?

This paper has presented new equivalence relations
involving multi-qubit systems. In particular, this formalism provides a new
way of simplifying the system-bath interactions in open quantum
systems.

\begin{acknowledgements}
We thank L. Quiroga and F. Rodr{\' i}guez for discussions. A.O.-C.
and C.F.L. thank the Clarendon Fund and University College
(Oxford) respectively, for financial support.
\end{acknowledgements}

\end{document}